\documentclass{emulateapj}

\newcommand{\mdot}{M$_{\odot}$\,}

\shorttitle{Photometry of 5 HVSs}
\shortauthors{Ginsburg et al.}

\begin{document}

\title{Variability of Hypervelocity Stars}

\author{Idan Ginsburg\altaffilmark{1} \altaffilmark{3}} \email{idan.ginsburg@dartmouth.edu}
\author{Warren R. Brown\altaffilmark{2}} \email{wbrown@cfa.harvard.edu} 
\author{Gary. A. Wegner\altaffilmark{1}} \email{gary.wegner@dartmouth.edu}

\altaffiltext{1}{Department of Physics and 
Astronomy, Dartmouth College, 6127 Wilder Laboratory, Hanover, NH 03755, USA}
\altaffiltext{2}{Smithsonian Astrophysical Observatory, 60 Garden St., Cambridge, MA 02138, US}
\altaffiltext{3}{Visiting Astronomer, Kitt Peak National Observatory, National Optical Astronomy Observatory, 
which is operated by the Association of Universities for Research in Astronomy (AURA) under cooperative agreement 
with the National Science Foundation.}

\begin{abstract}
We present time-series photometry of 11 hypervelocity stars (HVSs) to constrain their nature.
Known HVSs are mostly late-B spectral type objects that may be either main-sequence (MS) or evolved blue horizontal branch (BHB) stars. 
Fortunately, MS stars at these effective temperatures, $T_{eff} \sim$ 12,000 K, 
are good candidates for being a class of variable stars known as slowly pulsating B stars
(SPBs). We obtained photometry on four nights at the WIYN\footnote{The WIYN Observatory is a joint facility of the University of 
Wisconsin-Madison, Indiana University, Yale University, and the National Optical Astronomy Observatory.} 3.5 m telescope, and 
on six nights on the 2.4 m Hiltner telescope.
Using sinusoidal fits, 
we constrain four of our targets to have periods between $P \sim 0.2 - 2$ days, with a mean value of 0.6 days. 
Our amplitudes vary between $A = 0.5 - 3$\%.  
This suggests that these four HVSs are SPBs. We discuss a possible origin for these stars, and why further 
observations are necessary.

\end{abstract}

\keywords{Galaxy: center --- Galaxy: halo --- Galaxy: kinematics and dynamics --- stars: individual(SDSS J090744.99+024506.88, 
SDSS J091301.01+305119.83, SDSS J091759.47+672238.35, SDSS J113312.12+010824.87, SDSS J105248.30-000133.94)}

\section{Introduction}

\citet{Hills:88} 
theorized that a binary star system disrupted by a massive black hole (MBH) could result in the unbound ejection of 
one component as a HVS. \citet{Brown:05} discovered the first HVS
in the Galactic halo, and currently over 20 HVSs have been identified in the Milky Way
(\citealt{Edelmann:05}; \citealt{Hirsch:05}; \citealt{Brown:06a};
\citealt{Brown:06b}; \citealt{Brown:07b}; \citealt{Brown:09b}; \citealt{Brown:12}). 
Due to their large distances, the nature of the HVSs can be difficult to determine.
One useful technique is time-series photometry, however before this paper 
only HVS1 had been thus observed \citep[hereafter F06]{Fuentes:06}.
This paper present the results of time-series photometry for 11 HVSs.

The Milky Way houses Sgr A*, a massive black hole (MBH) of $\sim 4 \times 10^6$ M$_{\odot}$ 
(e.g. \citealt{Ghez:05}; \citealt{Ghez:08}; \citealt{Gillessen:2009}).
In Hill's scenario, HVSs are a natural consequence of a binary star system interacting with
this MBH. However, a number of different mechanisms 
have been proposed to produce HVSs,
including the inspiral of an intermediate-mass black hole (\citealt{Yu-Tremaine:03}; \citealt{Sesana:09}),
the disruption of a triple-star system (\citealt{Perets:09b}; \citealt{Ginsburg:3}), and interactions 
between stars and stellar-mass black holes \citep{Oleary-Loeb:08}. 
Observational and theoretical
evidence point to the Hill's mechanism as the most likely source for these HVSs
(e.g. \citealt{Ginsburg:1}; \citealt{Perets:09a}; \citealt{Brown:10b}; \citealt{Brown:12b}).
Simulations show that when a HVS is produced, the companion is left in a highly eccentric orbit around Sgr A*. 
This agrees with the known orbits of a number of stars observed orbiting within 1\arcsec\, of Sgr A* 
(e.g. \citealt{Scho:03}; \citealt{Ghez:05}),
thereby suggesting that some of the stars nearest Sgr A* (so-called S-stars), 
are former companions to HVSs \citep{Ginsburg:1}. Simulations further show that
a binary star system disrupted by the MBH may result in a collision between the two stars, and if the collisional velocity is
small enough the system may coalesce (\citealt{Ginsburg:2}; \citealt{Antonini:11}). HVSs may also be used 
to probe the shape of the Galactic halo \citep{Gnedin:05}, and they may even house planets \citep{Ginsburg:4}. 

Consequently, HVSs offer a wealth of knowledge, and it is important to understand their nature. One difficulty
is that HVSs may be MS or evolved BHB stars. Known HVSs are late-B spectral type objects,  
and MS and BHB stars at this T$_{eff} \sim12,000$ K have very similar surface gravities. Depending on their nature,
the intrinsic luminosity of an observed HVS may differ by a factor of $\sim 4$ and consequently the estimated distances to the HVS may differ
by a factor of $\sim 2$. The ages may also differ by as much as an order of magnitude. 
Using the present-day stellar mass function, \citet{Demarque-Virani:07} argue that HVSs are most likely
evolved low-mass stars. However, echelle spectroscopic observations indicate that HVS3 is a MS B star of M $\sim9$ \mdot (\citealt{Edelmann:05}; 
\citealt{Przybilla:08}; \citealt{Bonanos:08}) while HVS5, HVS7 and HVS8 are MS stars of M $\sim3.5$ \mdot each 
(\citealt{Brown:12b}; \citealt{Przybilla:08b}; \citealt{Lopez-Bonanos:08}). 

The remaining HVSs are too faint to be studied with echelle spectroscopy with existing telescopes.
However, the effective temperature of
HVSs makes them candidates for being SPBs.
SPBs were first introduced by \citet{Waelkens-91} who found seven intermediate B-type stars 
with photometric variations of a few millimagnitudes (mmag) and periods of $\sim 1$ day. 
This variability is due to $g$-mode pulsations, which are believed to be driven by the 
$\kappa$-mechanism (e.g. \citealt{Dziembowski:93}; \citealt{Gautschy-Saio:01}). 
Observed SPBs have periods 0.5-4 days,
spectral range between B2 and B9, masses 3-7 \mdot, and T$_{eff}$ = 12,000-18,000 K
(\citealt{Waelkens-91}; \citealt{Waelkens:98}; \citealt{Gautschy-Saio:02}; \citealt{DeCat:02}). 
Furthermore, all SPBs are observed to be 
slow rotators although why this is the case is not well understood \citep{Ushomirsky-Bildsten}.

BHB stars are bluewards of the RR Lyrae instability strip and are not observed to pulsate
(\citealt{Contreras:05}; \citealt{Catelan}). Consequently, 
the detection of mmag variability with a period of $\sim 1$ day is indicative of a SPB, and are
the two observable properties we are seeking.
F06 found a period for HVS1 consistent with that of a SPB. However, \citet{Turner-AN} argue that the 
observed decrease in brightness of HVS1 may be due to extinction.  

To check for variability we took time-series photometry of 11 HVSs over the course of four nights on the WIYN 3.5 m telescope 
and follow-up observations over the course of six more nights on the Hiltner 2.4 m telescope.
In \S 2 we discuss our observations. In \S 3 we discuss our analysis. We conclude our results in \S 4.

\section{Observations}
\subsection{WIYN Observations}

The first set of observations were taken the nights of 2012 February 23 -- 26 with the Mini-Mosaic Imager \citep{Saha-MIMO}
at the 3.5 m WIYN telescope at Kitt Peak National Observatory. All observations were in the SDSS {\it g}-band, and the results are summarized
in Table \ref{tab_short}. The Mini-Mosaic Imager has a field of view of 9.6\arcmin\, $\times$ 9.6\arcmin\, 
with 0.141 arcsec pixel$^{-1}$.
Our goal was to detect photometric variability to within a few
percent amplitude. In order to obtain enough photon statistics, faint objects such as HVS1 required
longer exposure times, up to 1200 s, while brighter objects such as HVS5 had exposure
times as short as 300 s. This provided high signal-to-noise ratio (S/N) $\sim 100 - 200$. 
We measured the photometry differentially using nearby stars 
of similar colors to within $\sim 0.6$ mag in $(g-r)$, identified by Sloan Digital Sky Survey photometry (SDSS; \citealt{York:SDSS}).
The raw images were reduced in IRAF\footnote{Imaging Reduction and Analysis Facilities (IRAF) is distributed by the National Optical
Astronomy Observatories which are operated by the Association of Universities for Research in Astronomy (AURA) under cooperative
agreement with the National Science Foundation.} 
(see \citealt{Tody}) using ccdproc. 
Photometry was analyzed using package DAOPHOT \citep{Stetson-PHOT} and SExtractor \citep{Bertin-Extract}. 

\subsection{MDM Observations}
The second set of observations were taken the nights of 2012 May 11 -- 16 with the 4K imager at the 
2.4 m Hiltner telescope at the MDM Observatory. All observations were in the Johnson
{\it B}-band, and the results are summarized in Table \ref{tab_short}.
Conversion between $B$ and $g$ is given in a number of papers (e.g. \citealt{Jester:05}; \citealt{Karaali:05}).
Since we chose the standards and targets to have similar colors to within $\sim 0.6$ mag in $(g-r)$,
any systematic differences will be minimal 
and have no affect on our overall results.
The 4K imager has a field of view of 21.3\arcmin\, $\times$ 21.3\arcmin\, with 0.315
arcsec pixel$^{-1}$. We concentrated on the five HVSs that based upon our first observations appeared to be variable 
(in Table \ref{tab_short}). The MDM data have significantly lower precision
since most of our targets set early in May and the higher airmasses led to poorer seeing. 
We reduced our 4K data using software pipeline developed by 
Jason Eastman\footnote{http://www.astronomy.ohio-state.edu/$\sim$jdeast/4k/proc4k.pro}. We then analyzed our photometry in the same
manner as that of our first set of observations.

Our errors are limited by photon statistics. Typical errors for our measurement with the WIYN are $\sim 0.5$\%. The Hiltner 
data are substantially poorer with errors $\sim 3$\% or larger. 
These large errors do not offer additional constraints on the HVSs, and thus are omitted from the calculations.
Nevertheless the Hiltner measurements do provide a consistency check for
our WIYN data, and we discuss them in more detail with regards to HVS5 and HVS7. We take 0.01 mag 
as the threshold for a significant detection of variability.

\begin{deluxetable}{ccrrrrrrrrcrl}
\tabletypesize{\scriptsize}
\tablecaption{Summary of Observations}
\tablewidth{0pt}
\tablehead{
\colhead{Star} & \colhead{WIYN Images} & \colhead{Variable?} & \colhead{MDM Images}
}
\startdata
HVS1 & 12 & Yes & 6\\
HVS4 & 10 & Yes & 6\\
HVS5 & 12 & Yes & 13\\
HVS6 & 8 & No & -\\
HVS7 & 8 & Yes & 12\\
HVS8 & 8 & No & -\\
HVS9 & 8 & No & -\\
HVS10 & 8 & No & -\\
HVS11 & 7 & No & -\\
HVS12 & 7 & No & -\\
HVS13 & 7 & Yes & 6\\
\hline
\enddata
\tablecomments{The leftmost column is the target, followed by
the number of exposures taken with the WIYN 3.5 m telescope (column 2), whether the data show any variability (column 3),
and the number of exposures taken in follow-up observations done with the Hiltner 2.4 m telescope (column 4).}
\label{tab_short}
\end{deluxetable}

%

\section{Results}

Of the 11 HVSs observed using the WIYN 3.5 m telescope, 
we find three HVSs with significant variability, one HVS with possible variability,
and HVS5 is ambiguous.
In all cases the period of variability is consistent with that of SPBs (see Table \ref{tab_outcome}). 
For each star, we fit our data to the model

\begin{equation}
y = A\,sin[2\pi ft_i + \phi] .
\end{equation}
where $A$ is the amplitude, $f$ the frequency, and $\phi$ our phase at time $t_i$. We consider periods in the range $3<P<120$ hours.
Periods shorter than three hours are not meaningful, since our observations for each HVS were taken no less than a few hours apart. Similarly,
the period must be less than five days since our observing program on the WIYN was a total of 4 nights. We constrain the amplitude
to be $0.005\leq A \leq 0.05$ which is consistent with our errors. We plot $\chi^2$ versus frequency for the five HVSs that showed
possible variability.

\begin{figure}[h]
\epsscale{2}
\plottwo{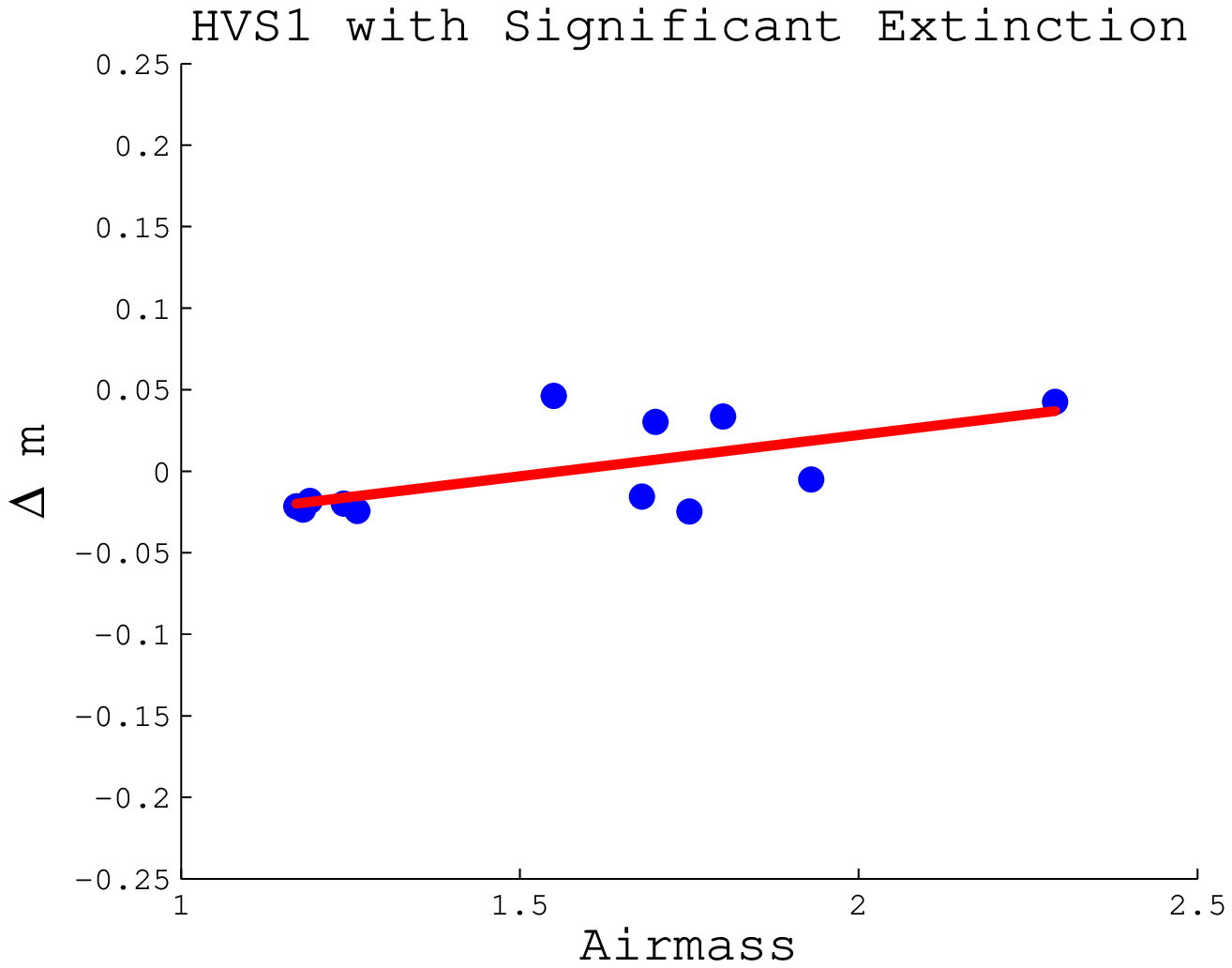}{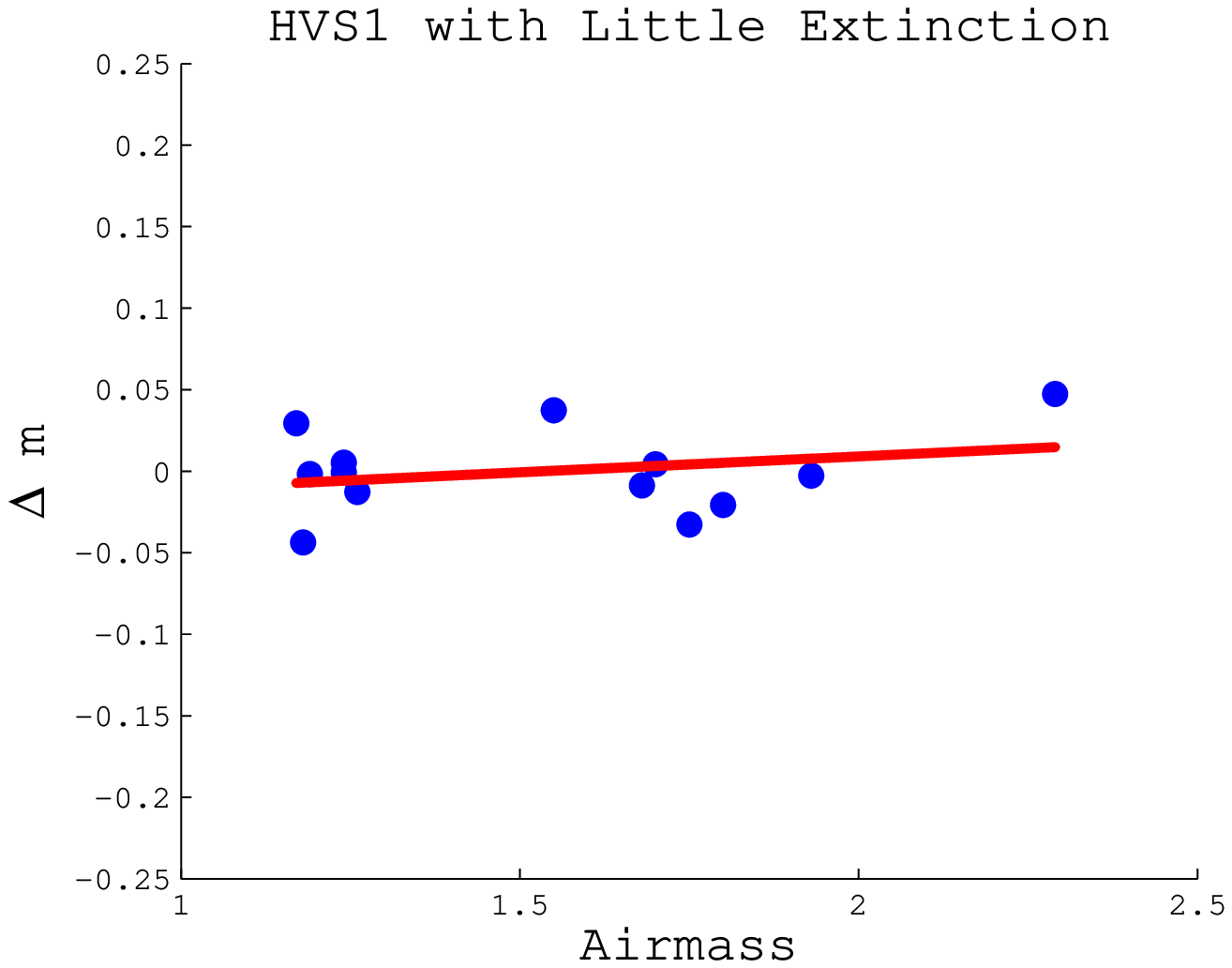}
\caption{The effect of airmass on observations is demonstrated here. 
Straight lines are best fits to the data. Note that HVS1 has $g-r$ = -0.2391.
Top: relative photometry of HVS1 versus airmass for a star with $g-r$ = 0.6907. 
Extinction produces a systematic difference in relative photometry of $\sim$ 0.05 mag.
Bottom: relative photometry of HVS1 versus airmass for a star with $g-r$ = 0.2673.
Extinction produces a systematic difference in relative photometry of $<$ 0.01 mag. Note that for 
our other targets, the systematic difference was significantly less than 0.01 mag. The color values were retrieved from 
SDSS ({http://skyserver.sdss3.org/DR8/en/tools/search/IQS.asp})}
\label{airmass}
\end{figure}

As noted by \citet{Turner-AN}, atmospheric extinction diminishes the brightness of blue stars more than that of red stars. 
For our differential photometry we choose stars with similar color to within $\sim 0.6$ mag in $(g-r)$ of our target HVSs. 
We examined the dependence of our observations on airmass and observed a nearly zero mag deviation in relative photometry for 
all targets except HVS1 which showed a very slight slope.
Consequently, a comparison star of redder color would produce a significant difference in relative photometry (see Figure \ref{airmass}). 


In order to determine the significance of our detections, we looked at a few goodness of fit tests.
The $\chi^2$ distribution is useful, but can be ambiguous. 
However, the well known $F$-test looks at two populations according to the $F$ distribution
given by
\begin{equation}
F = \frac{\chi^2_1/\nu_1}{\chi^2_2/\nu_2}
\end{equation}
where $f$ is our function and $\nu_1$ and $\nu_2$ are the degrees of freedom corresponding 
to $\chi^2_1$ and $\chi^2_2$ \citep{Bevington-Robinson}.
We calculate $\chi^2_1$ for $y=0$ and compare it with $\chi^2_2$ obtained by fitting our 
best fit values into equation 1.
Our results are summarized in Table \ref{tab_outcome}. 
In the following, we give notes on the individual stars. Error bars for the
period and amplitude were obtained using Monte Carlo methods and then calculating the RMS.


\subsection{HVS1}

\begin{figure}[h]
\epsscale{2}
\plottwo{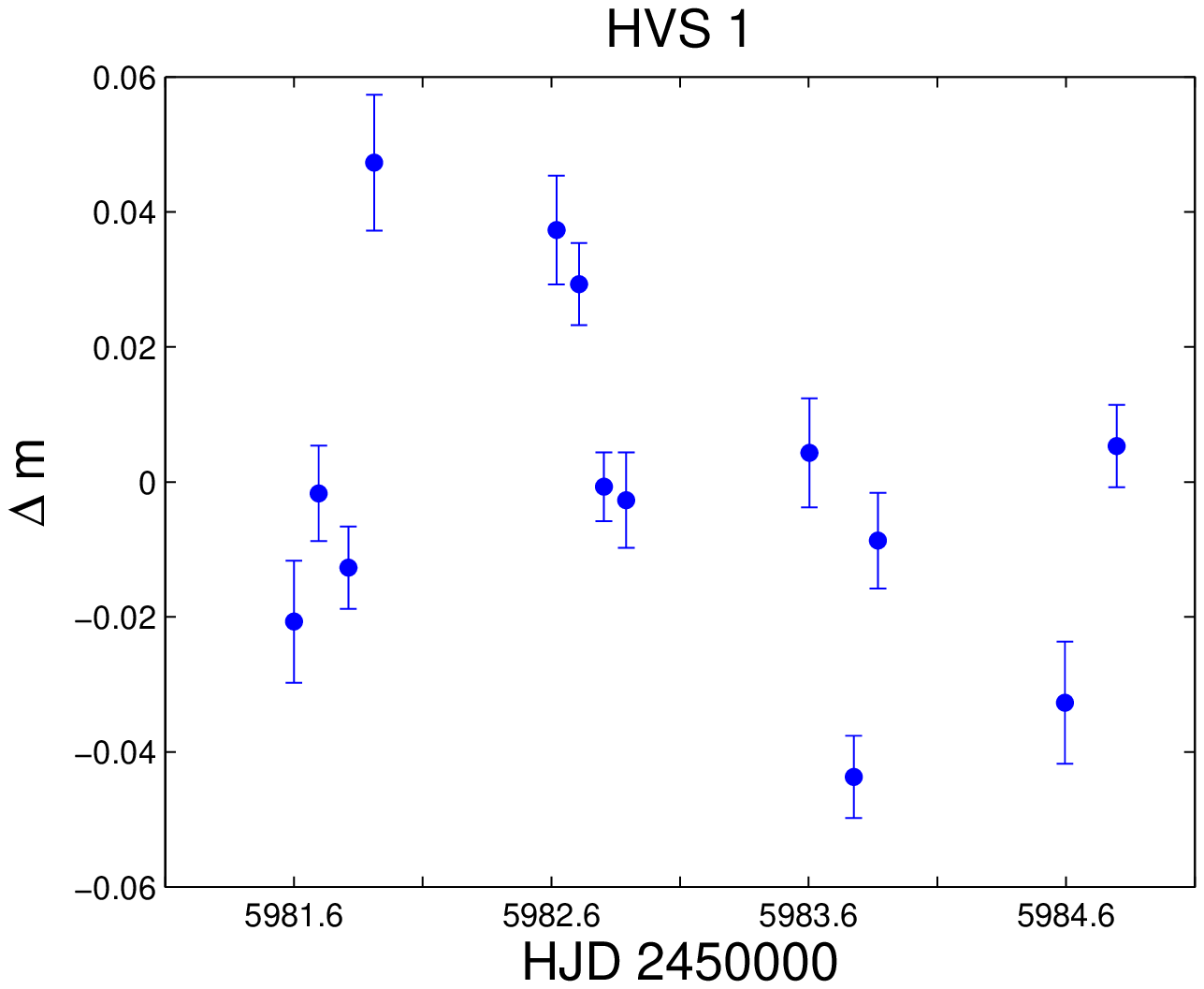}{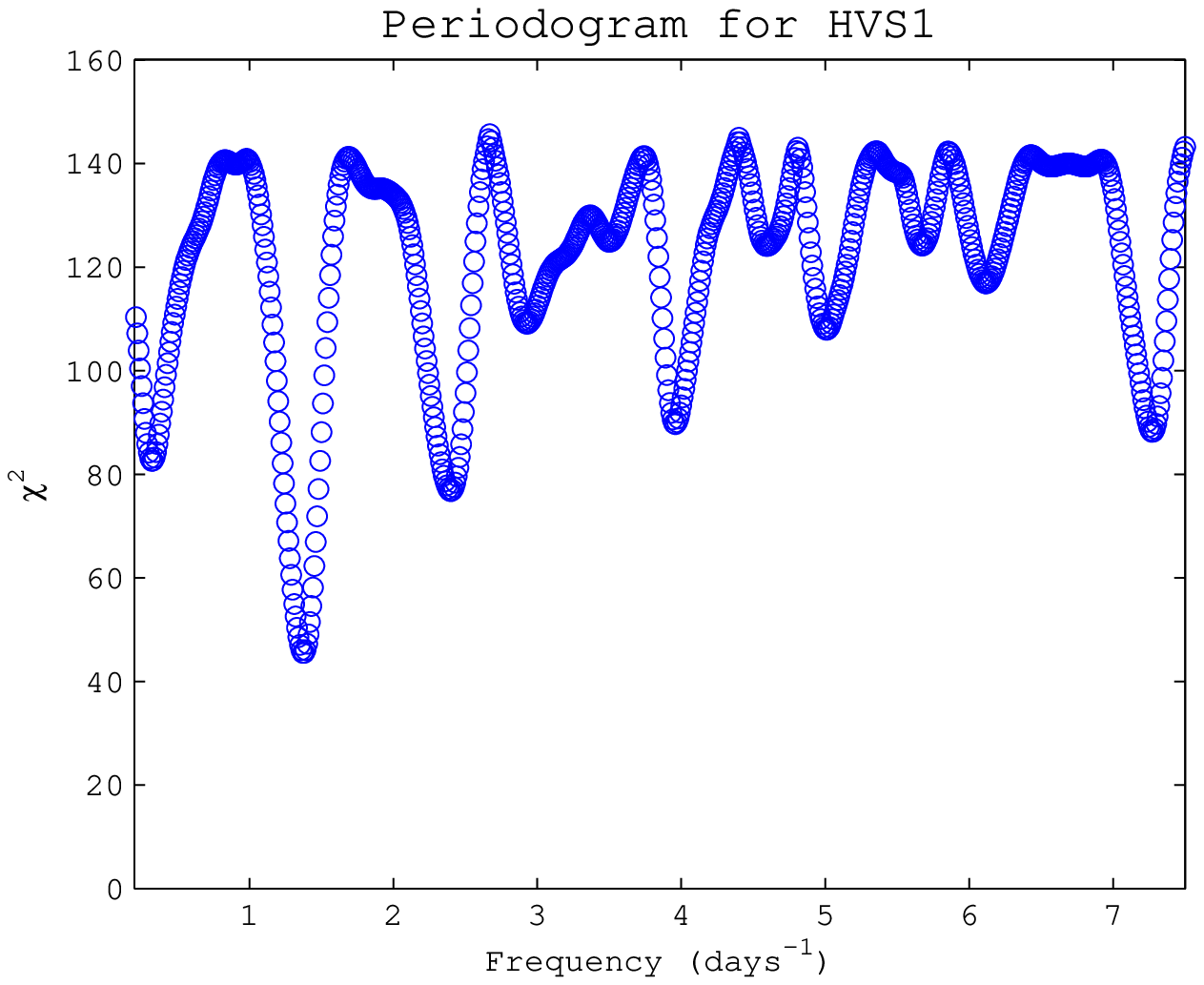}
\caption{Results for HVS1. Top: relative photometry of HVS1 with data taken from the WIYN 3.5 m telescope. For convenience 
the photometry was rescaled to have a mean of zero. This is $g$-band vs. HJD-2,450,000. Bottom: $\chi^2$ as a function
of period (days). The best-fit model gives $P = 0.72738 \pm 0.00767$ days.}
\label{HVS1}
\end{figure}

\begin{figure}[h]
\epsscale{1}
\plotone{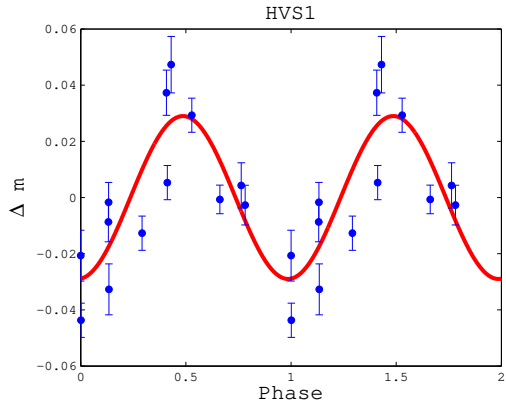}
\caption{Relative photometry of HVS1 as a function of phase angle. The light curve is folded according to the best-fit model with
$P = 0.72738$ days.}
\label{PHVS1}
\end{figure}

Before our measurements, HVS1 was the only HVS with time-series photometry. 
F06 carried out their observations over two nights with the 6.5 m telescope on the MMT followed
by four nights with the 1.2 m telescope at FLWO. They obtained both $g$ and $r$-band images. Although
they found no variability in the $r$-band, they did find significant variability in the $g$-band.
When comparing results, we find that our best fit amplitude of $A = 0.02878 \pm 0.00156$ mag agrees well with F06 who
found $A = 0.0280 \pm 0.0033$ mag.
If we calculate our differential photometry using a star with $g-r$ significantly larger than 
HVS1 (see the top panel of Figure \ref{airmass})
we find a best fit period of $P = 0.34567$ days which agrees to within 3\% with F06 who found $P = 0.355$ days. 
However, if atmospheric extinction is taken into account and we use a star with $g-r$ 
closer to that of HVS1 (see the bottom panel of Figure \ref{airmass}),
our most significant period is $P = 0.72738 \pm 0.00767$ days. 
Note that $P \sim 0.35$ days is our second most significant period given our data.
Figure \ref{HVS1} shows our relative photometry for HVS1 and our periodogram. 
Our $\chi^2_{min}$ = 45.5 for 10 degrees of freedom. 
Figure \ref{PHVS1} shows our best fit model with our WIYN data folded about our best fit period.
Our $F$-test resulted in a value of 0.0825 which has significance at the 1.6-sigma level. 

\subsection{HVS4}

\begin{figure}
\epsscale{2}
\plottwo{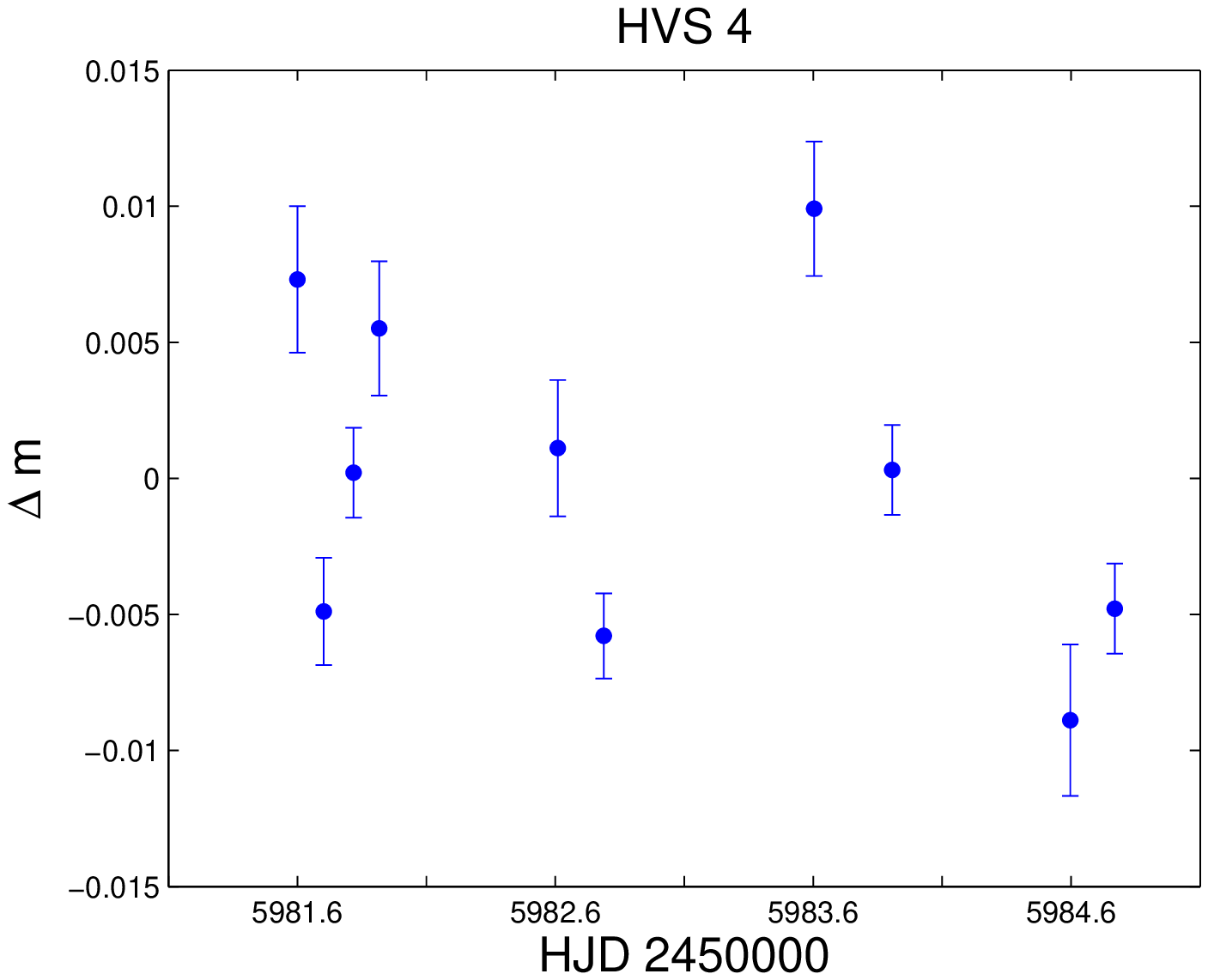}{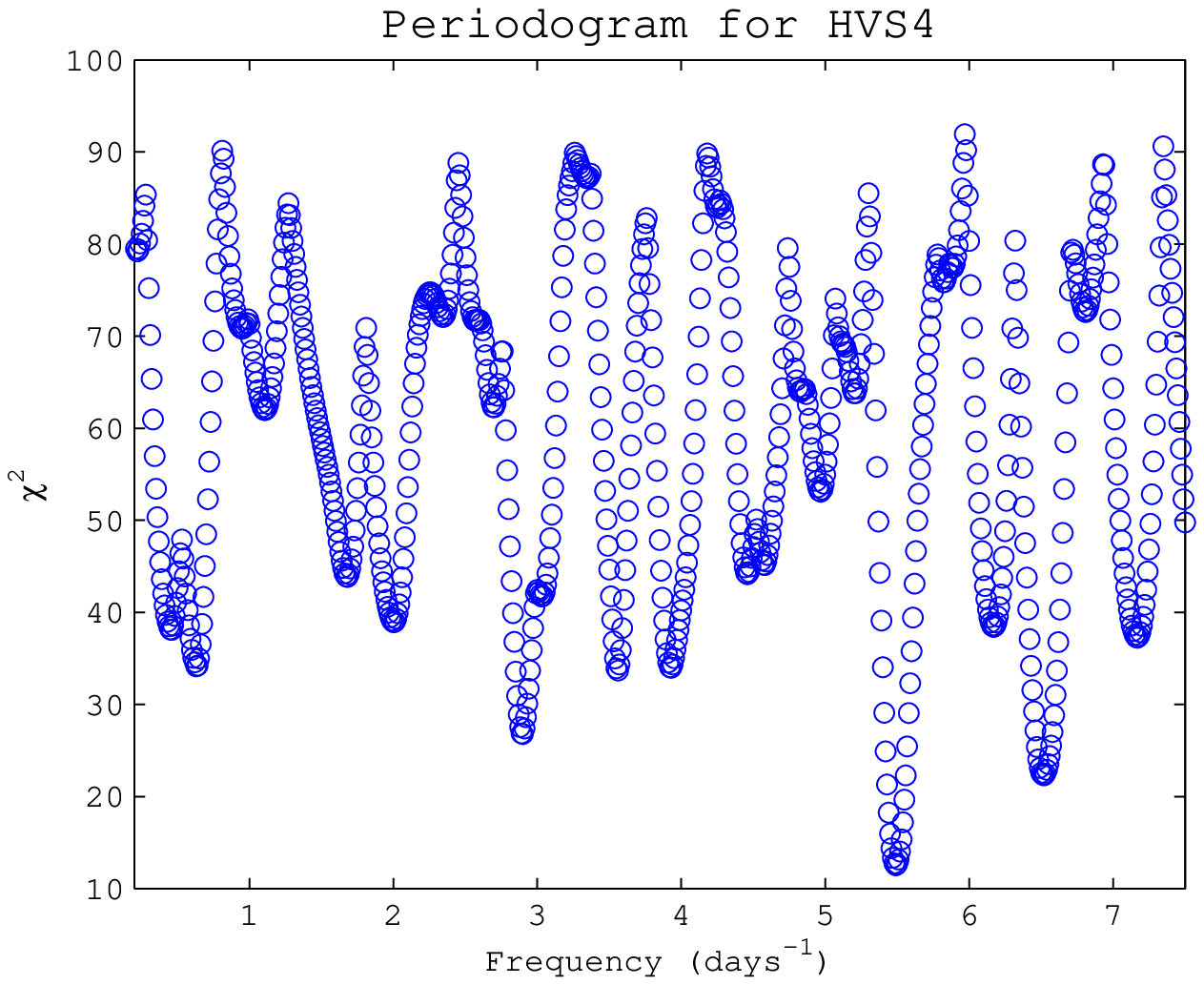}
\caption{Results for HVS4. Top: relative photometry of HVS4 with data taken from the WIYN 3.5 m telescope. For convenience 
the photometry was rescaled to have a mean of zero. This is $g$-band vs. HJD-2,450,000. Bottom: $\chi^2$ as a function
of period (days). The best-fit model gives $P = 0.18212 \pm 0.00057$ but we have aliases up to $P \sim 2$ days.}
\label{HVS4}
\end{figure}

\begin{figure}
\epsscale{1}
\plotone{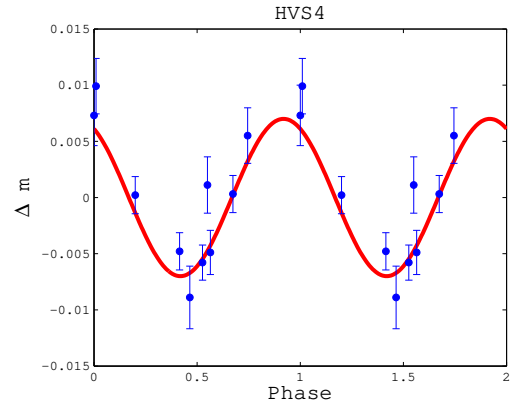}
\caption{Relative photometry of HVS4 as a function of phase angle. The light curve is folded according to the best-fit 
model with $P = 0.18212$ days.}
\label{PHVS4}
\end{figure}

We find a best fit period of $P = 0.18212 \pm 0.00057$ days. However, there are strong aliases ranging from $P \sim 0.15 - 2$ days.
Our best fit amplitude is $A$ = 0.00672 $\pm 0.00064$ mag. Our $\chi^2_{min}$ = 12.5 for 7 degrees of freedom.
Figure \ref{HVS4} shows our relative photometry for HVS4 and our periodogram. 
Figure \ref{PHVS4} shows our best fit model with our WIYN data folded about our best fit period.
Our $F$-test resulted in probability of 0.0463 which is significant at the two-sigma level.

\subsection{HVS5: An Ambiguous Object}

\begin{figure}
\epsscale{2}
\plottwo{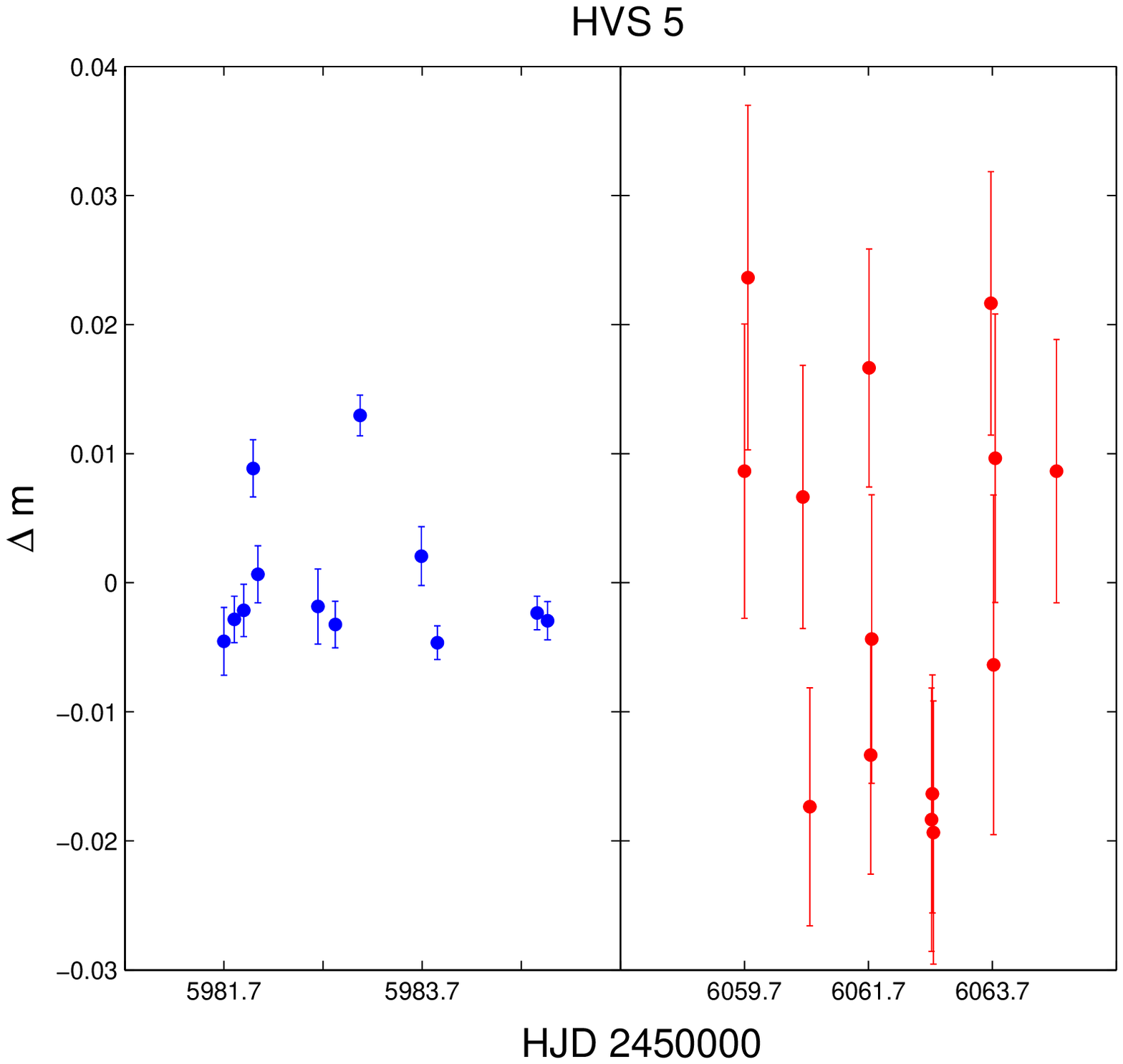}{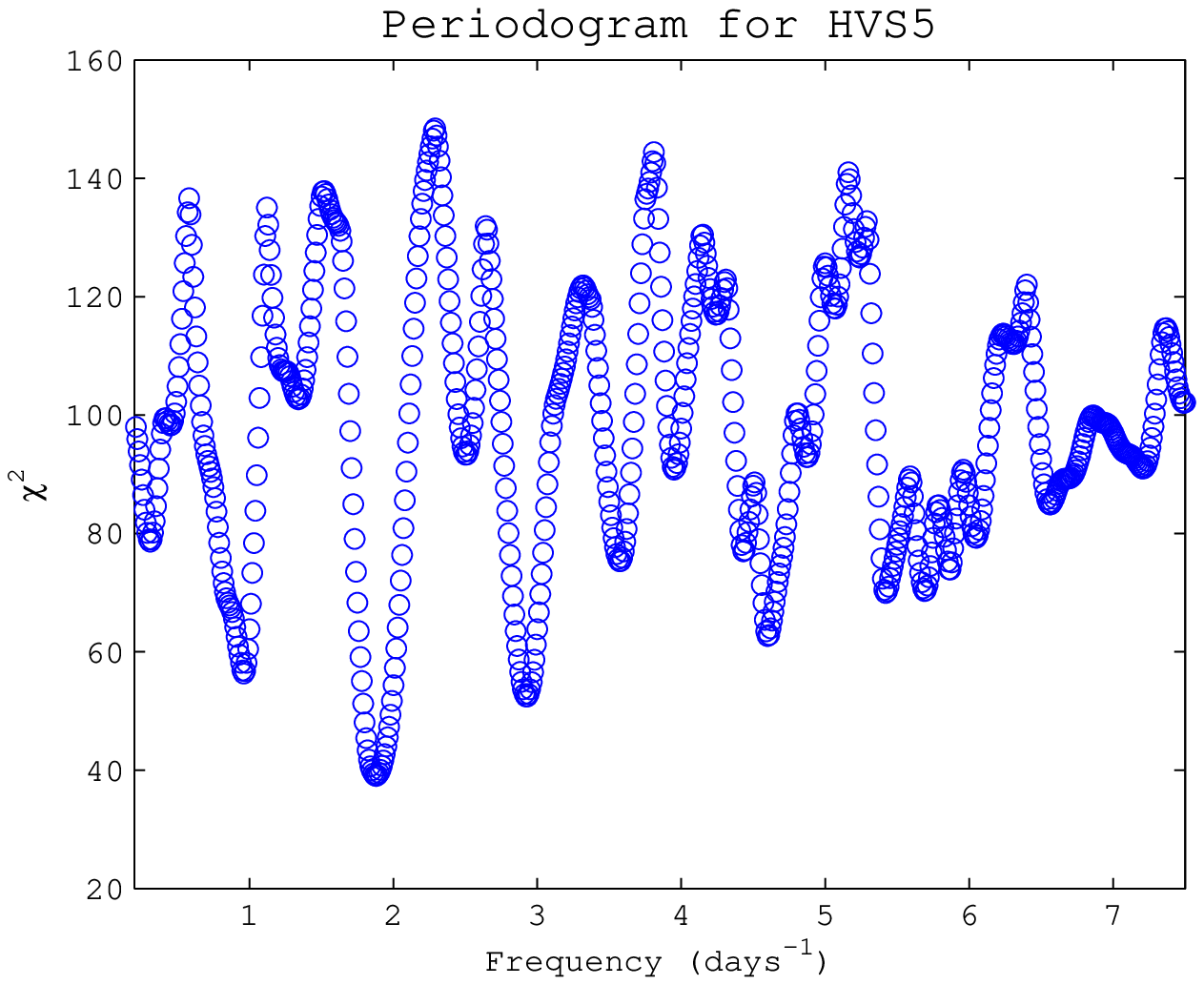}
\caption{Results for HVS5. Top: relative photometry of HVS5 with data taken from the WIYN 3.5 m (on the right) and the Hiltner 2.4 m 
telescope (on the left). For convenience 
the photometry was rescaled to have a mean of zero. This is $g$-band vs. HJD-2,450,000. Bottom: $\chi^2$ as a function
of period (days). The best-fit model gives $P = 0.53362 \pm 0.00666$.}
\label{HVS5}
\end{figure}

\begin{figure}
\epsscale{1}
\plotone{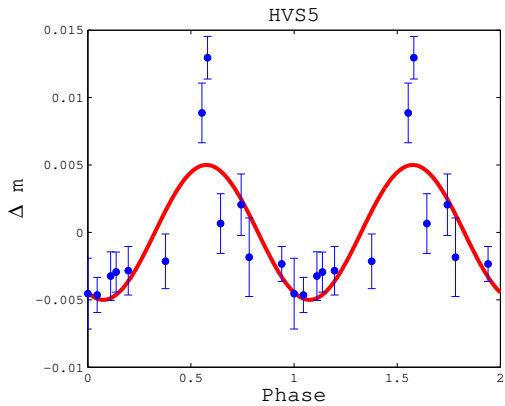}
\caption{Relative photometry of HVS5 as a function of phase angle. The light curve is folded according to 
the best-fit model with $P = 0.53362$ days.}
\label{PHVS5}
\end{figure}

HVS5 is the least constrained target. 
The variability suggests that HVS5 may be a SPB star, however
our $F$-test gave a probability of 0.1225 which is significant to only 1.2-sigma.
Therefore, we can not claim a detection for HVS5.  
Figure \ref{HVS5} shows our relative photometry for HVS5 and our periodogram. Our data from the Hiltner telescope 
correlated well with our data form the WIYN telescope, however the large errors ($\sim$ 2-3\%) do not offer any additional constraints 
on the variability of HVS5, and are only shown for completeness. 
We find a best fit period of $P = 0.53362 \pm 0.00666$ days, with strong aliases at $P = 0.337$ and 1.031 days. Our best fit amplitude is
$A$ = 0.00538 $\pm 0.00031$ mag. Our $\chi^2_{min}$ = 39.0 for 9 degrees of freedom. 
Figure \ref{PHVS5} shows our best fit model with our WIYN data folded about our best fit period.
Further observations our necessary to help determine the nature of HVS5.

\subsection{HVS7}

\begin{figure}
\epsscale{2}
\plottwo{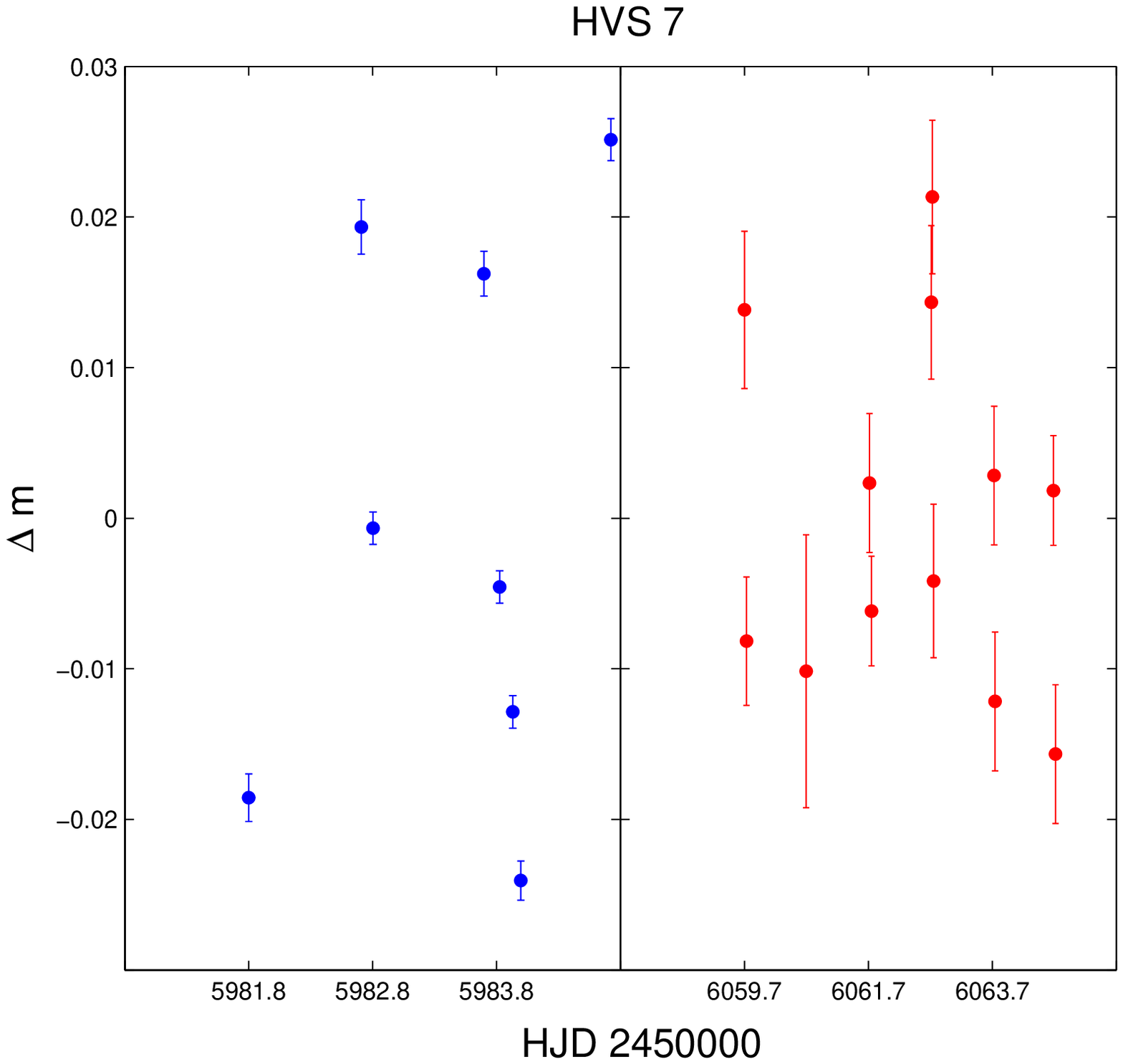}{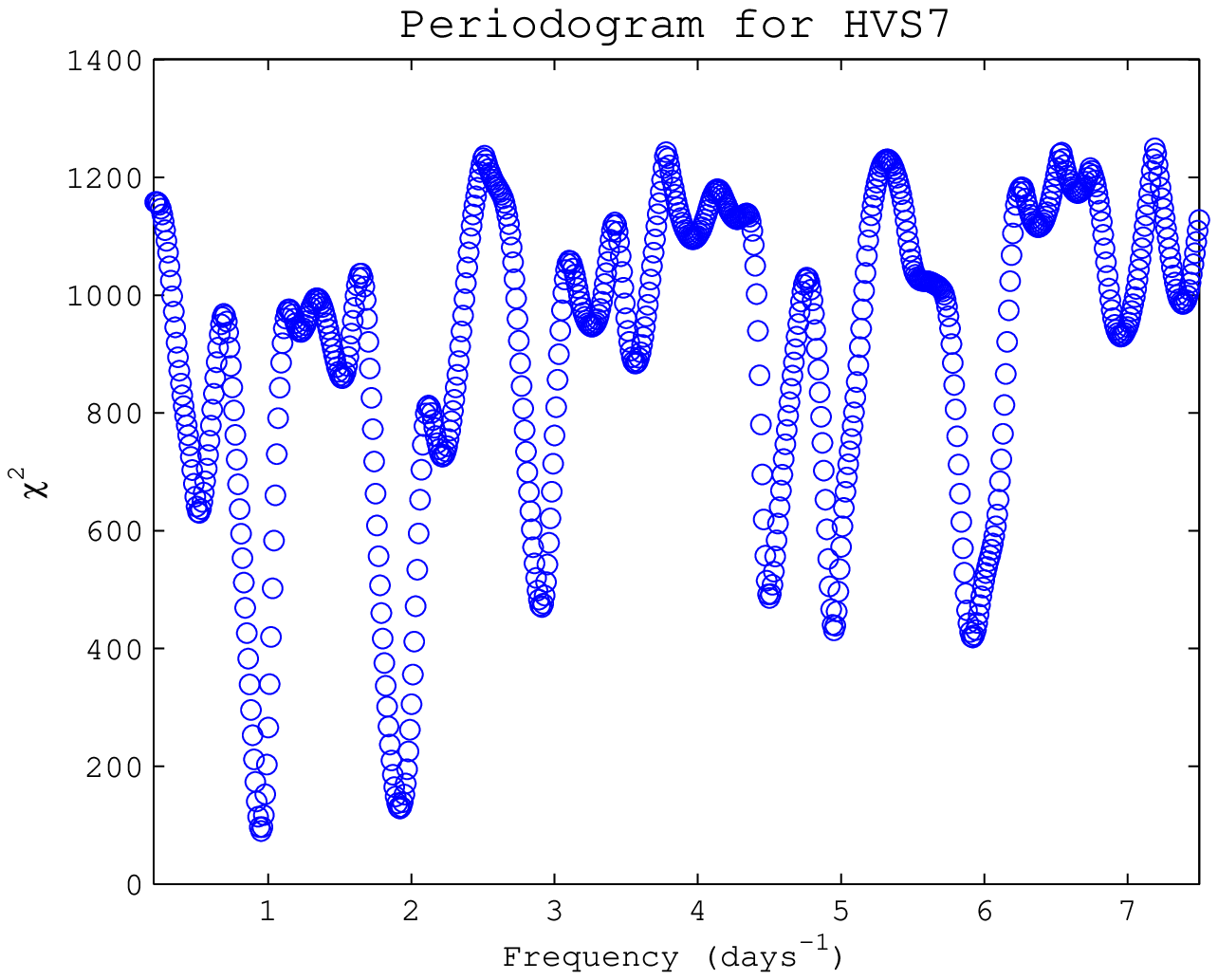}
\caption{Results for HVS7. Top: relative photometry of HVS7 with data taken from the WIYN 3.5 m (on the right) and the Hiltner 2.4 m 
telescope (on the left). For convenience 
the photometry was rescaled to have a mean of zero. This is $g$-band vs. HJD-2,450,000. Bottom: $\chi^2$ as a function
of period (days). The best-fit model gives $P = 1.05261 \pm 0.00194$ days.}
\label{HVS7}
\end{figure}

\begin{figure}
\epsscale{1}
\plotone{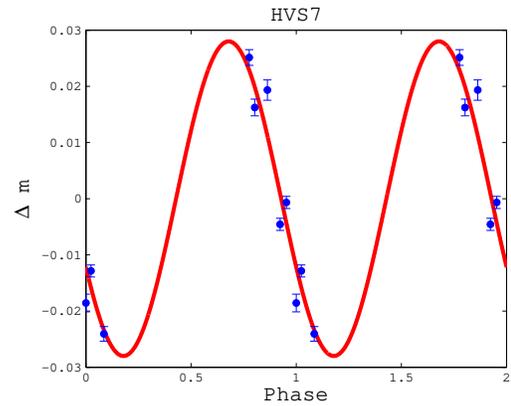}
\caption{Relative photometry of HVS7 as a function of phase angle. The light curve is folded according to the 
best-fit model with $P = 1.05261$ days.}
\label{PHVS7}
\end{figure}

HVS7 is our best constrained target. 
Our $F$-test gave a probability of 0.016 which has significance at the 2.5-sigma level.
We find a best fit period $P = 1.05261 \pm 0.00194$ days, with a strong alias at $P = 0.521$ days. Our best fit amplitude is
$A$ = 0.02812 $\pm 0.00056$ mag. Our $\chi^2_{min}$ = 90.2 for 5 degrees of freedom.
Figure \ref{HVS7} shows our relative photometry for HVS7 and our periodogram. 
Figure \ref{PHVS7} shows our best fit model with our WIYN data folded about our best fit period.
The data from the Hiltner telescope are shown for completeness.

%


\subsection{HVS13}

\begin{figure}
\epsscale{2}
\plottwo{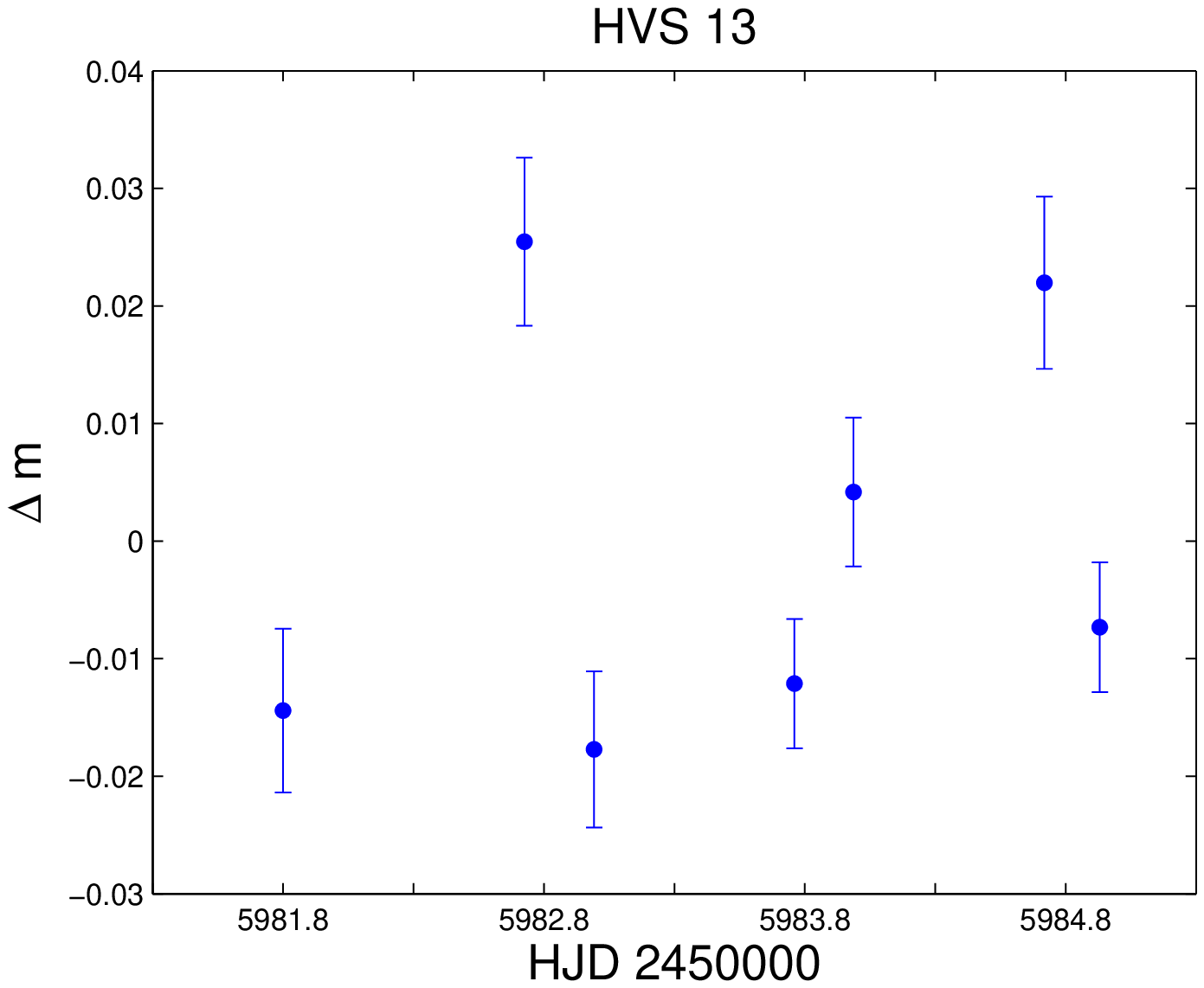}{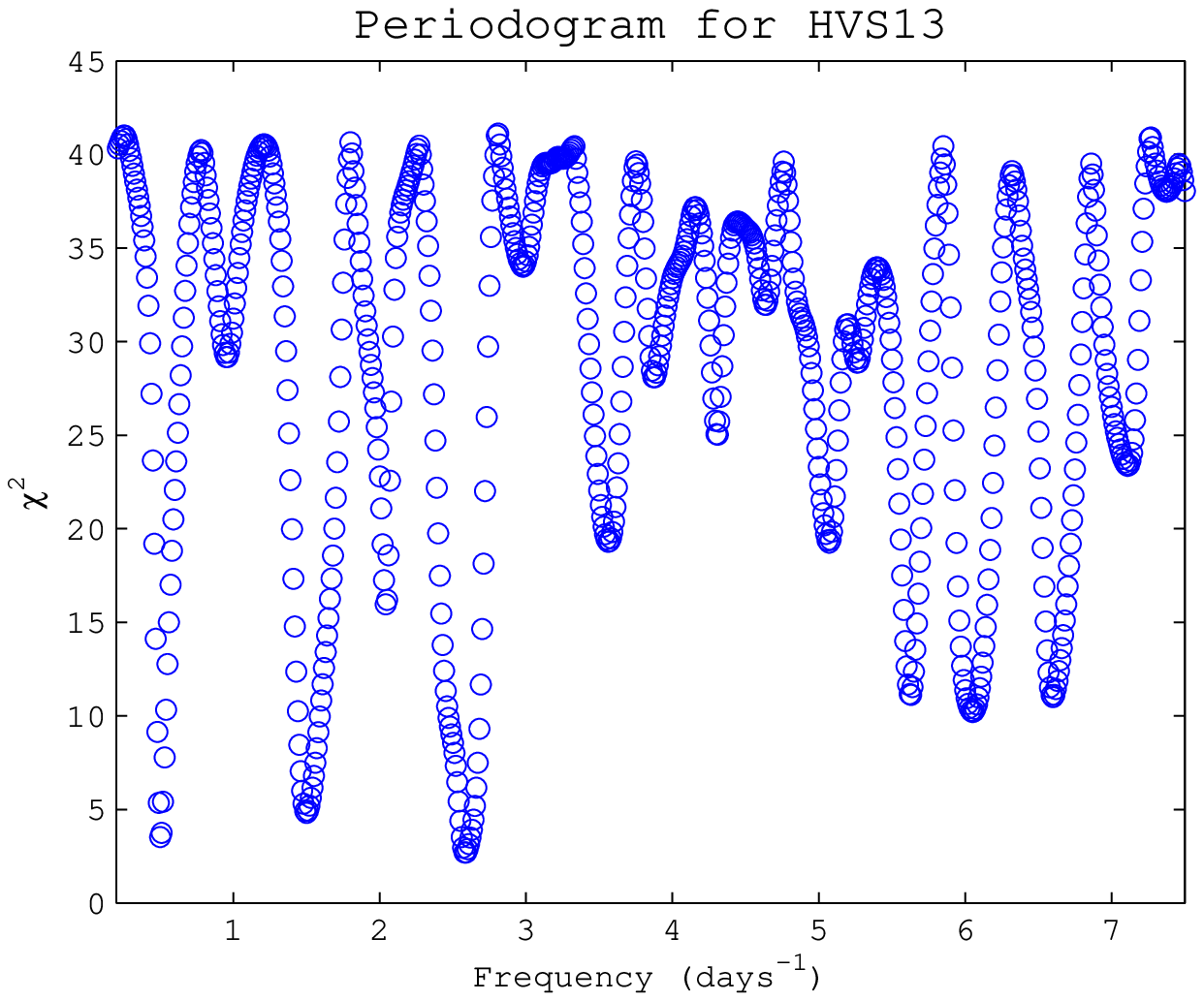}
\caption{Results for HVS13. Top: relative photometry of HVS13 with data taken from the WIYN 3.5 m telescope.  
For convenience the photometry was rescaled to have a mean of zero. This is $g$-band vs. HJD-2,450,000. Bottom: $\chi^2$ as a function
of period (days). The best-fit model gives $P = 0.38693 \pm 0.00402$ days.}
\label{HVS13}
\end{figure}

\begin{figure}
\epsscale{1}
\plotone{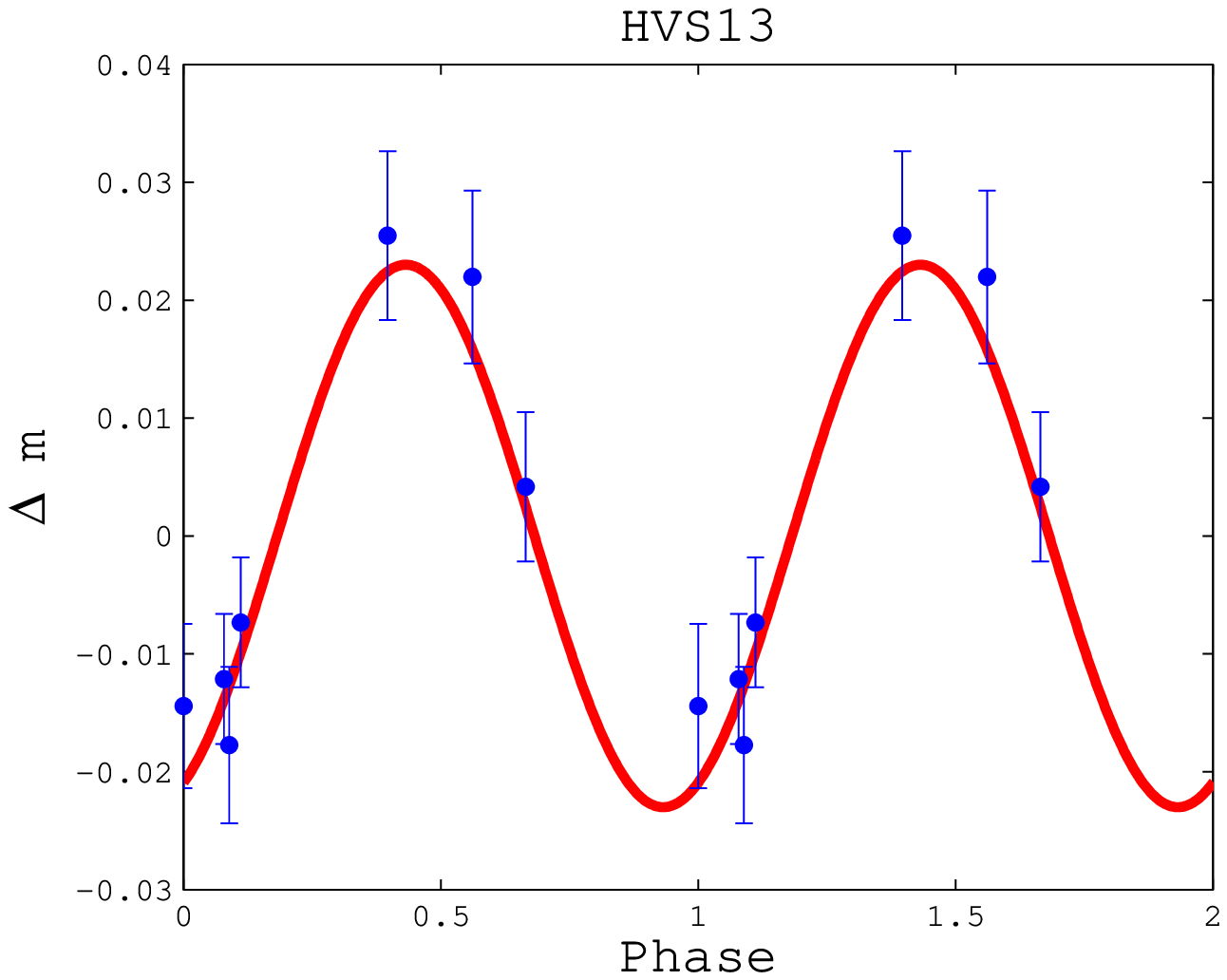}
\caption{Relative photometry of HVS13 as a function of phase angle. The light curve is folded according to the best-fit 
model with $P = 0.38693$ days.}
\label{PHVS13}
\end{figure}

HVS13 had the least amount of exposures, however our $F$-test gave a probability of 0.0283 which is significant to two-sigma.
Our $\chi^2_{min}$ = 2.71 for 4 degrees of freedom, which is the best value for all five targets.
We find the best fit period to be $P = 0.38693 \pm 0.00402$ days, with the strongest aliases at $P = 2.000$ and 0.667 days. The best fit amplitude is
$A$ = 0.02318 $\pm 0.00369$ mag. Figure \ref{HVS13} shows our relative photometry for HVS13 and our periodogram.
Figure \ref{PHVS13} shows our best fit model with our WIYN data folded about our best fit period.

\section{Summary and Conclusion}

\begin{deluxetable*}{ccrrrrrrrrcrl}[h*]
\tabletypesize{\scriptsize}
\tablecaption{Summary of Results for each SPB Candidate}
\tablewidth{0pt}
\tablehead{
\colhead{Star} & \colhead{RA} & \colhead{DEC} & \colhead{M$_B$} & \colhead{NOAO} &
\colhead{MDM} & \colhead{Period (days)} & \colhead{Amplitude (mag)} &
\colhead{$\chi^2$} & \colhead{P(F-test)}
}
\startdata
HVS1 & 9:07:44.993 & 2:45:06.88 &  19.687 & 12 & 6 & 0.72738$\pm$0.00767 & 0.02878$\pm$0.00156 & 45.5 & 0.0825\\
HVS4 & 9:13:01.011 & 30:51:19.83 & 18.314 & 10 & 6 & 0.18212$\pm$0.00057 & 0.00672$\pm$0.00064 & 12.5 & 0.0463 \\
HVS5 & 9:17:59.475 & 67:22:38.35 & 17.557 & 12 & 13 & 0.53362$\pm$0.00666 & 0.00538$\pm$0.00031 & 39.0 & 0.1225 \\
HVS7 & 11:33:12.123 & 1:08:24.87 & 17.637 & 8 & 12 & 1.05261$\pm$0.00194 & 0.02812$\pm$0.00056 & 90.2 & 0.0160 \\
HVS13 & 10:52:48.306 & -0:01:33.940 & 20.018 & 7 & 6 & 0.38693$\pm$0.00402 & 0.02318$\pm$0.00369 & 2.71 & 0.0283\\
\hline
\enddata
\tablecomments{The leftmost column is the name of the HVS. The next two columns are the 
right ascension (RA) and declination (DEC) respectively, and following is the absolute magnitude (M). Next is 
the number of images taken with the WIYN 3.5 m telescope (NOAO) and the Hiltner 2.4 m telescope (MDM) respectively. The best fit period is given, 
followed by the amplitude, and $\chi^2$. The rightmost column is the value from the $F$-test.}
\label{tab_outcome}
\end{deluxetable*}

We have taken time-series photometry of 11 HVSs (see Table \ref{tab_short}) and determined that
HVS1, HVS4, HVS7, and HVS13 show degrees of variability with best fit periods $\sim 0.2 - 1.0$ days and 
amplitudes $\sim 0.006-0.03$ mag. 
SPBs have observed periods 
between $0.5-4$ days which is consistent with our best fit periods.
The variability of the target HVSs are a few millimagnitudes which again is consistent for SPBs.
SPBs have masses $M_{\star} \sim 3 - 7$ M$_{\odot}$, and a number of confirmed SPBs (see \citealt{DeCat:02}) 
have mass $\sim 3$M$_{\odot}$ and T$_{eff} \sim 12,000$ K which agrees with the 
spectroscopically derived masses and temperatures of HVS1, HVS5, HVS7, and HVS8 \citep{Brown:12b}. 
Our $F$-test shows that HVS1 is suspect, with only a 1.6-sigma detection. HVS4 and HVS13 both
have a two-sigma detection, and HVS7 is detected at the 2.5-sigma level. HVS5 
was constrained at only the 1.2-sigma level, and thus can not be considered a detection, 
however it warrants further investigation.

Within 200 pc of the GC are regions dominated by massive Wolf-Rayet and OB supergiants (\citealt{Mauerhan:10}; \citealt{Dong:12}). 
However, at the innermost 0.05 pc the S-stars, young B-stars with masses $\sim 7 - 15$ \mdot dominate 
(\citealt{Ghez:03}; \citealt{Gillessen:2009a}). \citet{Ginsburg:1} suggested that 
the unexpected appearance of young stars around Sgr A* \citep{Ghez:03} can be explained at least in part by Hill's mechanism
where a binary star is disrupted by the MBH 
resulting in the production of a HVS of one component, while the other star falls into a highly eccentric orbit around Sgr A*. 
Defining ``arrival time'' as the time between its formation and ejection \citep{Brown:12b}, we find that
Hill's mechanism provides an arrival time of $t_{arr} \sim 0.1 - 1$ Gyr \citep{Merritt-Poon}
which is consistent with the expected lifetime of a MS B star of $\sim 3$ \mdot.
Furthermore, this scenario is supported by the results of $N$-body simulations (\citealt{Ginsburg:2}; \citealt{Antonini:10a}).
Recently, \citet{Bartko:10} found an isotropic distribution of B stars extending from the central arcsecond from Sgr A* to 12\arcsec. 
We identified four of 11 targets as as likely MS B stars. 
The fact that a significant percentage of our targets appear to be MS B stars helps support the case for an extended B star distribution.  

\begin{figure}
\epsscale{1}
\plotone{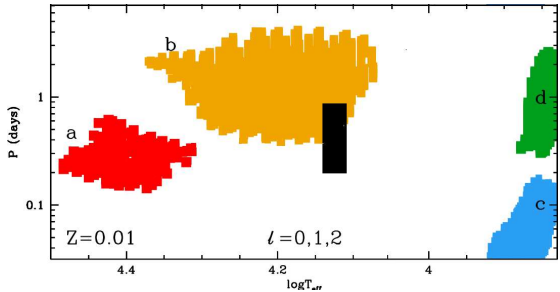}
\caption{Adapted from \citet{Degroote:09}. The theoretical instability domains of $l$ = 0,1,2 modes are represented in a 
log $T_{eff}$ versus log $P$ diagram. The regions are as follows: $a$ = $\beta$ Cep, $b$ = SPB,
$c$ = $\delta$ Sct, and $d$ = $\gamma$ Dor-type. These regions assume a metallicity Z = 0.01, however
results vary little for higher metallicity. Our results fall in the narrow black strip as shown. We assume a $T_{eff} \sim$ 12,000 K. }
\label{fig_finale}
\end{figure}

To date, only five of the known HVSs have been studied with high-resolution spectroscopy. HVS2 \citep{Hirsch:05} is believed to be
a subluminous O star, while the other four are MS B stars. 
HVS1 is the only HVS that has been observed with time-series photometry before our observations, and our results agree with 
F06 that HVS1 is a SPB. However, F06 derived a period of $P \sim 0.35$ days while our most 
significant period alias is twice this value.
HVS5 was recently observed with Keck HIRES spectroscopy \citep{Brown:12b}
which establishes it to be a MS B star, however our photometric 
data is ambiguous whether HVS5 may be a SPB star.
Further observations will be necessary in order to confirm the nature of HVS5.
HVS7 and HVS8 \citep{Lopez-Bonanos:08}
are both believed to be MS B stars. Our results for HVS7 show it to be a SPB with $P \sim 1.0$ days. We did not detect a variability for
HVS8. However, we can not rule out the possibility that it is a SPB with $P \geq 3$ days. The only other HVS with a detected variability
was HVS13 with a period between $P \sim 0.4 - 2$ days, suggesting it is a SPB. 
Without further observations, the remaining HVSs 
are either: SPBs with $P \geq 3$ days, SPBs with amplitudes below 0.01 mag, MS B stars but non-SPBs, or BHB stars. 
Figure \ref{fig_finale}, from \citealt{Degroote:09} 
summarizes our results in the context of the instability domain. The $x$-axis gives the period in days, and 
the $y$-axis denotes log of T$_{eff}$. 
Region $a$ shows the location of $\beta$ Cephei stars, 
which are early-type B stars (B0-B2.5) with variability of several hours \citep{Stankov-Handler}. 
$\delta$ Scuti stars, shown in region $c$, are of spectral type A and F with typical periods of 0.02-0.25 days \citep{Breger:07}.
Region $d$ shows $\gamma$ Doradus variables with periods similar to SPBs, $\sim 0.3-3$ days, however
they are of later spectral type A or F \citep{Pollard:09}. 
SPBs lie within region $b$.
Currently, all known HVSs on the MS are B type stars
and may be SPBs that lie within a small narrow strip of the instability domain illustrated by the black rectangle. 
However, further observations may support or modify the current paradigm.


\acknowledgments

We wish to thank the staff at NOAO and the WIYN telescope, especially Dianne Harmer and our
telescope operators Dave Summers and Karen Butler. We also thank the staff at 
the Hiltner telescope, in particular Bob Barr. 
We are grateful to Thiago Brito for all his assistance and discussions, and also Avi Loeb for
proofreading our draft. This research was supported in part by NOAO and Dartmouth College funds.

\bibliographystyle{apj.bst}
\bibliography{HVP.bib}

\end{document}